# The Variable O VIII Warm Absorber in MCG–6–30–15

Chiko OTANI[1], Tsuneo KII[2], Christopher S. REYNOLDS[3], Andrew C. FABIAN[3],
Kazushi IWASAWA[3,4], Kiyoshi HAYASHIDA[5], Hajime INOUE[2], Hideyo KUNIEDA[4],
Fumiyoshi MAKINO[2], Masaru MATSUOKA[1], and Yasuo TANAKA[6]

[1] *The Institute of Physical and Chemical Research (RIKEN), Hirosawa, Wako, Saitama 351-01*
*(E-mail (CO) otani@postman.riken.go.jp)*
[2] *Institute of the Space and Astronautical Science, Yoshinodai, Sagamihara, Kanagawa 229*
[3] *Institute of Astronomy, Madingley Road, Cambridge CB3 OHA, UK*
[4] *Department of Astrophysics, Nagoya University, Furo-cho, Chikusa-ku, Nagoya 464-01*
[5] *Department of Earth and Space Science, Osaka University, Machikaneyama, Toyonaka, Osaka 560*
[6] *Max-Planck-Institut für Extraterrestrische Physik, Giessenbachstrasse, D-85740 Garching, Germany*

**Abstract**

We present the results of a 4 day ASCA observation of the Seyfert galaxy MCG–6–30–15, focusing on the nature of the X-ray absorption by the warm absorber, characterized by the K-edges of the highly ionized oxygen, O VII and O VIII. We confirm that the column density of O VIII changes on a timescale of $\sim 10^4$ s when the X-ray continuum flux decreases. The significant anti-correlation of column density with continuum flux gives direct evidence that the warm absorber is photoionized by the X-ray continuum. From the timescale of the variation of the O VIII column density, we estimate that it originates from gas within a radius of about $10^{17}$ cm of the central engine. In contrast, the depth of the O VII edge shows no response to the continuum flux, which indicates that it originates in gas at larger radii. Our results strongly suggest that there are two warm absorbing regions; one located near or within the Broad Line Region, the other associated with the outer molecular torus, scattering medium or Narrow Line Region.

## 1. Introduction

Soft X-ray observations of many Seyfert 1 galaxies have now shown evidence for an absorption edge which is due to partially-ionized oxygen along the line of sight to the active nucleus (Nandra & Pounds 1992; Nandra et al. 1993; Fabian et al. 1994; Mihara et al. 1994). Since the absorbing medium is plausibly at about $10^5$ K, it is known as the warm absorber. Its high occurrence indicates that the fraction of the nucleus covered by this material is high, perhaps 50 per cent, which, combined with a typical column density of $\sim 10^{22}\,\mathrm{cm}^{-2}$, suggests that it is an important constituent of the region surrounding the nucleus. The product of covering fraction and column density are at least as large as that for the clouds responsible for the optical/UV broad line emission.

The location of the warm absorber has been unclear. If it is assumed that it is in photoionization equilibrium, then it must lie within about 30 pc of the nucleus. This value is deduced from the ionization parameter $\xi = L/nR^2 \approx 30\,\mathrm{erg\,cm\,s^{-1}}$ obtained from fits to the data in MCG–6–30–15 (Fabian et al 1994), where $L$ is the X-ray luminosity ($\sim 10^{43}\,\mathrm{erg\,s^{-1}}$), and the product $nR$ is the maximum column density ($\sim 10^{22}\,\mathrm{cm}^{-2}$).

Further progress can be made from considerations of the variability of the warm absorber. ASCA data from 1993 show that the warm absorber in MCG–6–30–15 varies on timescales of a few hours (Reynolds et al. 1995). This provides a lower limit on the recombination timescale of the gas and thus an upper limit on the radius, which was less than $\sim 10^{17}$ cm. The situation with those data was confusing however, since it appeared that the ionization parameter did not scale simply and linearly with ionizing luminosity, as would be expected from the nature of that parameter, but that it varied in some more complex manner.

Here we report results from a 4 day ASCA observation of MCG–6–30–15 taken in 1994. The warm absorber changed in the last day, giving a deeper feature as the flux diminished. We show that this and previous observations can now be understood if there are *two* warm absorbers, instead of one. The outer one dominates the OVII edge and has a long recombination time so does not change, the inner one dominates the OVIII edge and has a re-combination time of $10^4$ s or less, so responds directly to the ionizing radiation.

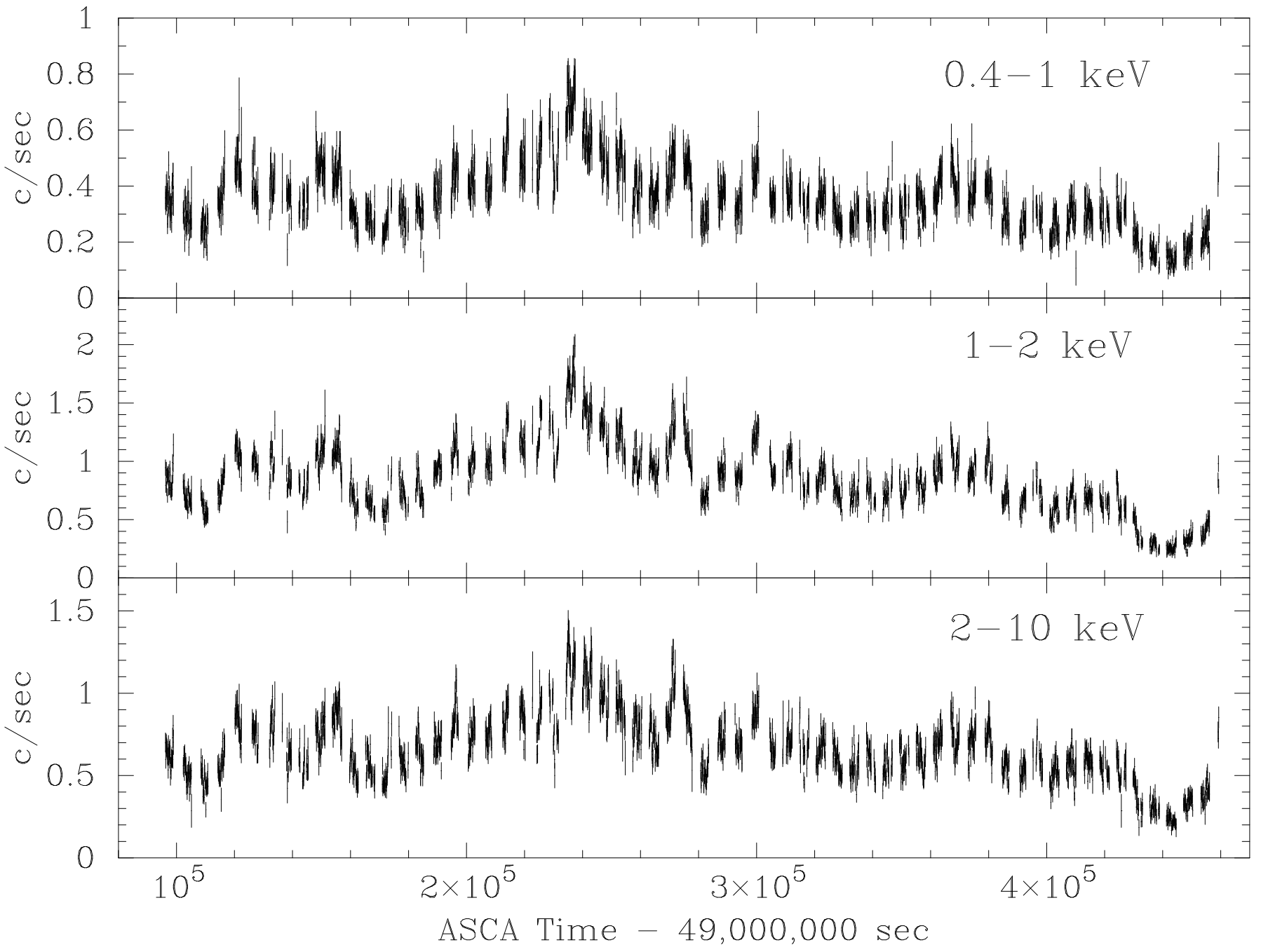

Fig. 1. SIS light curve of MCG–6–30–15 in three bands, 0.4–1 keV, 1–2 keV, and 2–10 keV. The horizontal axis is written in seconds from ASCA time = 49,000,000 s = 03:06:38 22 July 1994 (UT). The vertical axis is in units of count rate per SIS. Each bin width is typically 100 s. No background subtraction is applied because it is negligible.

The inner warm absorber is very thin, which suggests that it consists of many clouds. There may then be problems with the stability of the clouds, which presumably are pressure confined in some way, since there is only a narrow range of parameter space in which the necessary clouds can exist if the confining pressure is the thermal pressure of a hotter, more tenuous, medium also in photoionization equilibrium with the ionizing radiation (Reynolds & Fabian 1995). We explore the constraints imposed by these clouds and note that the inner warm absorber must contain a considerable reservoir of gas in which oxygen is completely ionized in order to provide O VIII when the ionizing flux drops.

## 2. Observation and Results

### *2.1. Mean Spectrum*

MCG–6–30–15 was observed with ASCA during 1994 July 23–27. Both Solid state Imaging Spectrometers, SIS-0(S0) and SIS-1(S1) were operated in 1-CCD Faint mode, and both Gas Imaging Spectrometers GIS-2(G2) and GIS-3(G3) were in normal PH mode. Since the GIS do not have the energy range to investigate the oxygen absorption edge features well, we concentrate here on the SIS data only.

All data used were selected from intervals of high and medium bit rate. The SIS data were selected using the following local criteria; a) times were more than 3 min from the SIS Radiation Belt Monitor flag being triggered by the South Atlantic Anomaly (SAA), b) the angle between the field of view and the edge of the bright and dark earth exceeded 25 and 5 degree, respectively, and c) the cutoff rigidity was greater than 4 GeV/c. After these selections, we also deleted data if d) the background rate was abnormally high near the SAA, e) there were any spurious events and f) the dark frame error was abnormal. The net exposure time was about 166 ksec in each SIS. The source varied during our observation by a factor~7 (figure 1).

Fitting the mean spectrum by a simple power-law model with a power-law model clearly shows the presence of edge-like absorption features between 0.7 and 2 keV, a complex excess around 5–7 keV, and further excess at 0.6 keV. The 0.7–2 keV feature is similar to that reported by Fabian et al. (1994). The line feature around 5–7 keV is probably due to reflection from the relativistic inner accretion disk as discussed by Tanaka et al. (1995).

In order to estimate the properties of the feature below

Table 1. Spectral fittings of 0.4–5 keV mean spectrum.

| Model | photon index $\Gamma$ | $N_{\rm H}$ ($10^{20}$ cm$^{-2}$) | $E_{\rm edge}$ (keV) | $\tau_{\rm edge}$ | $E^{\dagger}_{\rm line}$ (keV) | EW (eV) | $\chi^2/\nu$ |
|---|---|---|---|---|---|---|---|
| PL & 2 edges | $1.840^{+0.007}_{-0.008}$ | — | $0.720^{+0.005}_{-0.005}$ | $0.58^{+0.04}_{-0.04}$ | — | — | 994.6/513 |
| | | | $0.851^{+0.013}_{-0.013}$ | $0.20^{+0.03}_{-0.07}$ | | | |
| PL & 2 edges | $1.92^{+0.008}_{-0.009}$ | — | $0.721^{+0.004}_{-0.005}$ | $0.55^{+0.04}_{-0.04}$ | $0.61^{+0.01}_{-0.01}$ | $14^{+3}_{-3}$ | 905.2/511 |
| & 0.6 keV line | | | $0.849^{+0.015}_{-0.014}$ | $0.19^{+0.03}_{-0.03}$ | | | |
| PL & 2 edges | $1.93^{+0.02}_{-0.02}$ | $2.0^{+0.4}_{-0.4}$ | $0.720^{+0.004}_{-0.004}$ | $0.63^{+0.04}_{-0.04}$ | — | — | 896.4/512 |
| absorption | | | $0.864^{+0.013}_{-0.012}$ | $0.25^{+0.03}_{-0.04}$ | | | |
| PL & 2 edges | $1.92^{+0.02}_{-0.02}$ | $2.3^{+0.4}_{-0.4}$ | $0.721^{+0.005}_{-0.005}$ | $0.60^{+0.04}_{-0.03}$ | $0.60^{+0.01}_{-0.01}$ | $17^{+3}_{-4}$ | 780.3/510 |
| & line & absorption | | | $0.863^{+0.014}_{-0.012}$ | $0.24^{+0.03}_{-0.03}$ | | | |

† Line width is fixed to zero. Line energy is not corrected for redshift.

1 keV, we now restrict the energy range of our analysis to below 4 keV, because of the iron line feature above 5 keV. At least two absorption edges are needed in order to explain the broad trough around 0.7–1 keV. In addition, we included the excess neutral absorption above the Galactic Hydrogen column density ($N_{\rm H} = 4.06 \times 10^{20}$ cm$^{-2}$; Elvis, Lockman & Wilkes 1989) and a Gaussian line at 0.6 keV to suppress the hollow at 0.5 keV and the excess at 0.6 keV. It is plausible that these features are of instrumental origin, because the change of the spectral slope as seen below may cause such a feature near these energy where the response function has the sharp feature. In addition, the response matrix for these energies still includes some uncertainties due to the absence of the good calibrator for SIS. The final fitting results are shown in table 1. (Errors indicate 90% confidence levels for 2 interesting parameters.) The fluctuation of the derived parameters for the absorption edges are almost consistent with each other, which means that the inclusion of the absorption and a 0.6 keV line do not change our results on the edge parameters. The absorption edge model is only affected by the local structure at and above the edge threshold energy.

The rest-frame threshold energy of the lower edge is determined as $0.721 \pm 0.005$ keV which is near, but slightly lower than, that of the O VII edge. This tendency is almost unchanged by the models used and is also seen in the fitting results of spectra divided in time, as shown below. Although the threshold energy of the higher energy edge depends slightly on the model, its range is consistent with the O VIII edge energy, 0.871 keV. (Although the energies are similar to those of neutral iron L and neon K edges, the required high abundances of these elements, and lack of iron M and neon L absorption, indicate that this is mere coincidence.) We conclude that the observed absorption feature is dominated by O VII and O VIII edges, as shown by Fabian et al. (1994). We emphasize that the small energy shift is greater in O VII than in O VIII. The hydrogen-equivalent column densities of observed O VII and O VIII are $N_{\rm H}$(O VII)= $(3.0^{+0.2}_{-0.2}) \times 10^{21}$ cm$^{-2}$ and $N_{\rm H}$(O VIII)= $(2.8^{+0.4}_{-1.0}) \times 10^{21}$ cm$^{-2}$, respectively, assuming cosmic abundance for the absorbing matter. The mean column density is consistent with that in the first observation of Fabian et al. (1994).

We also tried a simple warm absorber model as described in Fabian et al. (1994), allowing the photon index to vary. The parameters obtained are $\xi = 17^{+2}_{-2}$ erg cm s$^{-1}$ and $N_{\rm H} = (4.6^{+0.3}_{-0.2}) \times 10^{21}$ cm$^{-2}$, when the extra neutral absorption and the 0.6 keV line are included. The model is worse than the two-edge model due mainly to the energy shift of the O VII edge. We will mainly use the two-edge model below.

### 2.2. *Time Variability of Warm Absorber*

In order to follow the change in the low-energy absorption features, we divided the whole 4-day observation into 17 parts, each of which has about 10 ks integration time. The background for each spectrum is subtracted using the spectrum obtained in off-source regions from the same CCD chip and the same time interval. (Note that the background subtraction has little effect on our results.) The observation log for these divisions is shown in table 2. We fit spectra using the model which contains a power-law continuum, two edges, and neutral absorption above the Galactic value. The results are listed in table 3 and shown in figure 2. (Errors indicate 90% confidence levels for 2 interesting parameters.) There is a clear dependence of the maximum optical depth of O VIII absorption edge with the incident continuum flux. The response of the edge is especially clear at the end of the observation. The difference of spectra in epochs 14 and 17 are shown in figure 3. The correlation with count rate and O VIII edge depth is expressed by a power-law of count rate :

$$\tau_{\rm OVIII} = (0.38 \pm 0.05) C^{-(0.96 \pm 0.30)} \qquad (1)$$

where $C$ is the count rate. 90% errors are quoted for interesting one parameter.

Table 2. Log for divided spectra.

| No. | Epoch | | Exp. | Flux |
|---|---|---|---|---|
| | start | end | (sec) | (c/s/SIS) |
| 1 | 7/23 05:49 | 7/23 11:30 | 10500 | 1.22 |
| 2 | 7/23 12:28 | 7/23 19:30 | 9500 | 1.62 |
| 3 | 7/23 20:11 | 7/24 00:18 | 9500 | 1.64 |
| 4 | 7/24 00:59 | 7/24 05:05 | 8000 | 1.17 |
| 5 | 7/24 05:46 | 7/24 11:29 | 10500 | 1.60 |
| 6 | 7/24 12:28 | 7/24 19:28 | 10300 | 1.94 |
| 7 | 7/24 20:09 | 7/25 00:16 | 9300 | 2.48 |
| 8 | 7/25 00:57 | 7/25 05:04 | 9100 | 1.78 |
| 9 | 7/25 05:44 | 7/25 09:51 | 9000 | 1.76 |
| 10 | 7/25 10:43 | 7/25 16:59 | 9300 | 1.63 |
| 11 | 7/25 17:17 | 7/25 22:39 | 10600 | 1.43 |
| 12 | 7/25 23:19 | 7/26 05:02 | 9700 | 1.23 |
| 13 | 7/26 05:42 | 7/26 09:50 | 9200 | 1.55 |
| 14 | 7/26 10:40 | 7/26 16:56 | 9800 | 1.39 |
| 15 | 7/26 17:12 | 7/26 22:38 | 11400 | 1.16 |
| 16 | 7/26 23:17 | 7/27 05:01 | 10400 | 0.89 |
| 17 | 7/27 05:41 | 7/27 10:43 | 9800 | 0.65 |

In contrast, the depth of the O VII absorption edge is consistent with being constant during our observation. These tendencies of both edges are quantitatively consistent with those observed in the PV phase (Fabian et al. 1994; Reynolds et al. 1995), in which the optical depth of O VII and O VIII were $\tau_{\rm OVII} = 0.53 \pm 0.07$ and $\tau_{\rm OVIII} = 0.19 \pm 0.05$ in the observation on 1993 July 8–9 ( the count rate was 1.4 cts $s^{-1}$ in SIS-0 ), and $\tau_{\rm OVII} = 0.63 \pm 0.08$ and $\tau_{\rm OVIII} = 0.44 \pm 0.07$ in the observation on 1993 July 31 – August 1 (the count rate was then 0.9 cts $s^{-1}$ in SIS-0 ). A transient change in the depth of the O VIII edge also occurred in the 1993 July data (Reynolds et al. 1995), similar to the event found here in the 1994 data. The recovery of the O VIII depth in the 1994 data is also suggested just after the minimum of the flux at epoch 4 and 12 in our light curve, but is not significant statistically.

## 3. Discussion

### *3.1. Constraints on the location of the absorbing material*

The current observations provide the best constraints yet on the state and location of the warm absorbing material in this bright Seyfert 1 galaxy. It is clearly seen that the variability of the warm absorber is characterized by a variable O VIII edge and a constant O VII edge. More precisely, the O VII edge appears to be constant over timescales of months to years whereas the O VIII edge can dramatically increase in optical depth over timescales of $10^4$ s or so. The depth of the O VIII edge is anti-correlated with the primary ionizing flux, thereby providing direct evidence that the material is photoionized.

If the O VII and O VIII edges arise in the same material, (the so-called one-zone models) the recombination and photoionization timescales of these two dominant ions would be comparable. Thus, an unknown mechanism would have to be invoked to stabilize the ionization fraction of O VII whilst allowing the O VIII to respond to the changing ionizing flux. This failure of simple one-zone models prompts us to consider models in which the O VII and O VIII edges originate from spatially distinct regions, one of which is steady and one of which undergoes the observed variations (the two-zone model). Another indication that there are two absorbing regions comes from the possible systematic velocity difference between the two absorption edges discussed in the previous section. Assuming that photoionization controls the ionization state of the warm material, we are led to consider the following model: O VII ions in the region responsible for the O VII edge have a long recombination timescale (i.e. weeks or more) whereas the O VIII edge arises from more highly ionized material (in which most oxygen is fully stripped) in which the O IX ions have a recombination timescale of $10^4$ s or less. A decrease in primary ionizing flux is then accompanied by recombination of O IX to O VIII giving the observed variation of the O VIII edge depth. Figure 4 shows the ionization fractions of various states of oxygen as a function of $\xi$. We postulate that the O VIII absorber exists in the right-hand portion of this figure in which the fraction of O VIII, $f_{\rm O8}$, is a decreasing function of $\xi$. In this regime (where oxygen is predominantly in the fully ionized state) $f_{\rm O8} \propto F^{-1}$ where F is the ionizing flux. To see this, note that balancing photoionization rates with recombination rates gives

$$f_{\rm O8} \sigma_{\rm O8} F = f_{\rm O9} n_{\rm e} \alpha \qquad (2)$$

where $\sigma_{\rm O8}$ is a frequency weighted mean photoionization cross section for O VIII, and $\alpha$ is the recombination co-efficient for O IX. Using $f_{O9} \sim 1$ gives $f_{O8} \propto F^{-1}$. The observed correlation between the O VIII edge depth and the SIS count rate (figure 2b) is in good agreement with this, thereby providing quantitative evidence that photoionization dominates the physics of this plasma.

Table 3. Fitting results of time-divided SIS spectra with a power-law plus two-edge model. The neutral absorption is also included.

| No. | count rate (c/s/SIS) | photon index $\Gamma$ | $N_{\rm H}$ ($10^{20}$cm$^{-2}$) | $E_{\rm O\,VII}$ (keV) | $\tau_{\rm O\,VII}$ | $E_{\rm O\,VIII}$ (keV) | $\tau_{\rm O\,VIII}$ | $\chi^2$/d.o.f. |
|---|---|---|---|---|---|---|---|---|
| 1 | 1.22 | $1.95^{+0.07}_{-0.07}$ | $0.9^{+1.6}_{-0.9}$ | $0.704^{+0.024}_{-0.030}$ | $0.56^{+0.16}_{-0.24}$ | $0.83^{+0.06}_{-0.07}$ | $0.27^{+0.25}_{-0.16}$ | 457.3/442 |
| 2 | 1.62 | $1.97^{+0.07}_{-0.06}$ | $2.7^{+1.5}_{-1.5}$ | $0.722^{+0.016}_{-0.017}$ | $0.61^{+0.12}_{-0.13}$ | $0.88^{+0.05}_{-0.06}$ | $0.19^{+0.11}_{-0.11}$ | 439.3/467 |
| 3 | 1.64 | $1.93^{+0.07}_{-0.07}$ | $1.9^{+1.8}_{-1.5}$ | $0.740^{+0.018}_{-0.034}$ | $0.61^{+0.16}_{-0.37}$ | $0.90^{+0.17}_{-0.12}$ | $0.15^{+0.29}_{-0.10}$ | 475.1/477 |
| 4 | 1.17 | $1.88^{+0.09}_{-0.08}$ | $1.1^{+2.1}_{-1.1}$ | $0.705^{+0.019}_{-0.020}$ | $0.62^{+0.18}_{-0.18}$ | $0.83^{+0.05}_{-0.03}$ | $0.38^{+0.17}_{-0.17}$ | 394.5/397 |
| 5 | 1.60 | $2.02^{+0.05}_{-0.06}$ | $4.2^{+1.5}_{-1.3}$ | $0.724^{+0.016}_{-0.016}$ | $0.66^{+0.21}_{-0.16}$ | $0.85^{+0.11}_{-0.04}$ | $0.30^{+0.16}_{-0.15}$ | 517.0/497 |
| 6 | 1.94 | $2.04^{+0.06}_{-0.06}$ | $3.7^{+1.3}_{-1.3}$ | $0.721^{+0.013}_{-0.015}$ | $0.62^{+0.13}_{-0.16}$ | $0.85^{+0.07}_{-0.06}$ | $0.18^{+0.16}_{-0.11}$ | 584.5/507 |
| 7 | 2.48 | $2.05^{+0.05}_{-0.06}$ | $4.5^{+1.2}_{-1.2}$ | $0.728^{+0.010}_{-0.011}$ | $0.77^{+0.11}_{-0.11}$ | $0.89^{+0.08}_{-0.04}$ | $0.20^{+0.09}_{-0.08}$ | 502.1/527 |
| 8 | 1.78 | $1.95^{+0.06}_{-0.05}$ | $2.8^{+1.4}_{-1.1}$ | $0.719^{+0.016}_{-0.023}$ | $0.69^{+0.13}_{-0.20}$ | $0.86^{+0.05}_{-0.07}$ | $0.22^{+0.19}_{-0.12}$ | 470.8/502 |
| 9 | 1.76 | $2.01^{+0.06}_{-0.07}$ | $5.2^{+1.6}_{-1.5}$ | $0.710^{+0.016}_{-0.016}$ | $0.70^{+0.11}_{-0.14}$ | $0.90^{+0.05}_{-0.06}$ | $0.24^{+0.10}_{-0.09}$ | 505.0/477 |
| 10 | 1.63 | $2.05^{+0.06}_{-0.06}$ | $5.7^{+1.5}_{-1.2}$ | $0.718^{+0.014}_{-0.014}$ | $0.68^{+0.14}_{-0.12}$ | $0.86^{+0.04}_{-0.04}$ | $0.24^{+0.12}_{-0.12}$ | 529.3/482 |
| 11 | 1.43 | $2.00^{+0.07}_{-0.06}$ | $4.1^{+1.5}_{-1.5}$ | $0.722^{+0.016}_{-0.018}$ | $0.67^{+0.15}_{-0.21}$ | $0.84^{+0.05}_{-0.05}$ | $0.24^{+0.18}_{-0.14}$ | 488.7/482 |
| 12 | 1.23 | $1.95^{+0.07}_{-0.07}$ | $2.4^{+1.6}_{-1.6}$ | $0.718^{+0.019}_{-0.019}$ | $0.52^{+0.16}_{-0.15}$ | $0.85^{+0.03}_{-0.03}$ | $0.37^{+0.13}_{-0.14}$ | 526.6/447 |
| 13 | 1.55 | $2.02^{+0.08}_{-0.07}$ | $3.5^{+1.7}_{-1.6}$ | $0.713^{+0.014}_{-0.014}$ | $0.67^{+0.12}_{-0.11}$ | $0.92^{+0.03}_{-0.04}$ | $0.25^{+0.09}_{-0.09}$ | 416.5/457 |
| 14 | 1.39 | $1.93^{+0.07}_{-0.07}$ | $1.9^{+1.4}_{-1.5}$ | $0.718^{+0.017}_{-0.019}$ | $0.64^{+0.14}_{-0.17}$ | $0.86^{+0.06}_{-0.06}$ | $0.21^{+0.14}_{-0.12}$ | 450.4/457 |
| 15 | 1.16 | $1.80^{+0.05}_{-0.03}$ | $0.0^{+1.4}_{-0.0}$ | $0.731^{+0.012}_{-0.013}$ | $0.54^{+0.16}_{-0.18}$ | $0.86^{+0.05}_{-0.04}$ | $0.35^{+0.16}_{-0.15}$ | 449.0/467 |
| 16 | 0.89 | $1.87^{+0.04}_{-0.04}$ | $0.0^{+0.5}_{-0.0}$ | $0.728^{+0.020}_{-0.022}$ | $0.64^{+0.14}_{-0.15}$ | $0.91^{+0.05}_{-0.04}$ | $0.34^{+0.13}_{-0.12}$ | 423.7/402 |
| 17 | 0.65 | $1.91^{+0.04}_{-0.05}$ | $0.0^{+0.6}_{-0.0}$ | $0.716^{+0.020}_{-0.019}$ | $0.62^{+0.18}_{-0.16}$ | $0.92^{+0.03}_{-0.03}$ | $0.63^{+0.14}_{-0.13}$ | 362.2/341 |

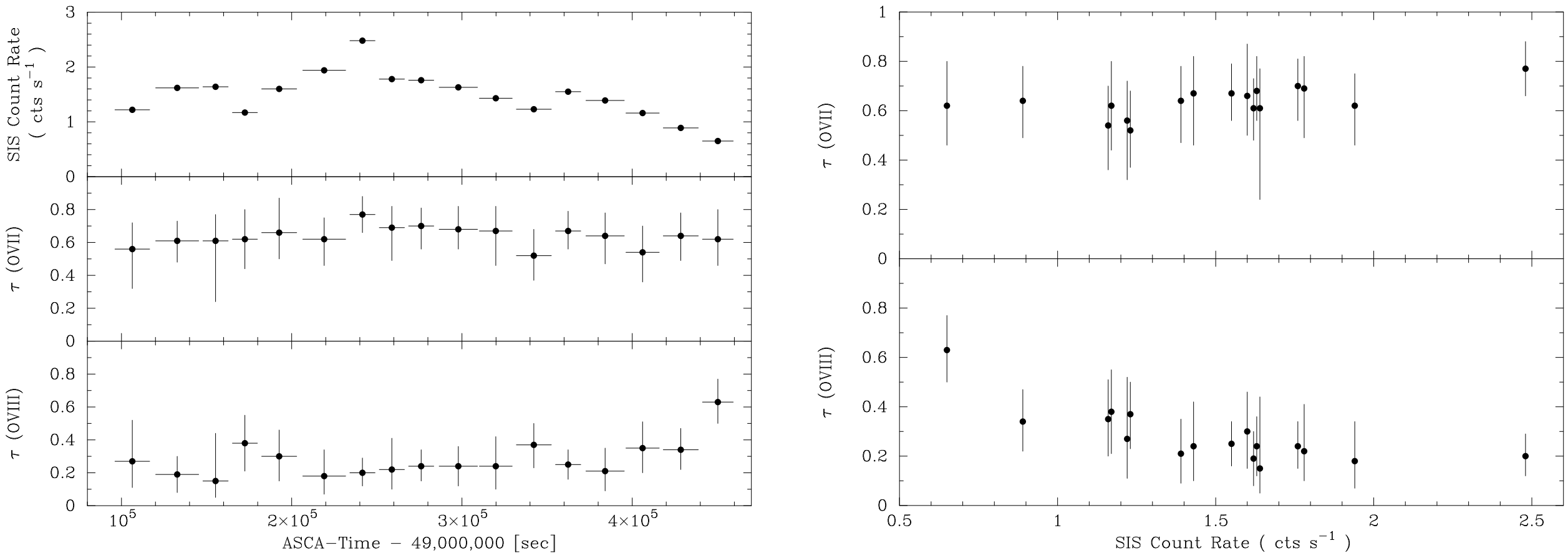

(a) (b)

Fig. 2. Time variability of the absorption edge features. (a) The light-curves of the continuum flux in count rate in 0.4–10 keV and the depths of the O VII and O VIII absorption edges. Typical duration of each point is 20,000 s. (b) The flux dependence of the absorption depths of O VII and O VIII edges.

To observationally characterize the two regions, we select two periods from the long *ASCA* observation. During the first 300 ks of the observation (which had the primary flux in a high state) the warm absorber is in the low O VIII state: a one-zone photoionization model based on CLOUDY (Ferland 1991) was fitted to this period of data and taken to represent the physical state of the O VII absorber. The best fit model has a column density of $N_{\rm W} = 4.6 \times 10^{21}\,{\rm cm}^{-2}$ and an ionization parameter of $\xi = 17.4\,{\rm erg\,cm\,s}^{-1}$. In contrast, the last 60 ks of the observation (with the primary flux in a low state) had the warm absorber in the high O VIII state. The above parameters for the O VII absorber were held fixed and an additional one-zone model was included to represent the O VIII absorber. This had the parameters $N_{\rm W} = 1.3 \times 10^{22}\,{\rm cm}^{-2}$ and $\xi = 74\,{\rm erg\,cm\,s}^{-1}$.

We examine the constraints on each of these regions in the ($R$, $\Delta R/R$) plane where $R$ is the distance of the ionized material from the central source of ionizing radiation and $\Delta R$ is the line of sight distance through the ionized material (which equates to the thickness of a shell in the thin-shell approximation). Physically, $\Delta R/R$ is representative of the volume filling factor the of material along the line of sight. Using the definition of the ionization parameter,

$$\xi = \frac{L}{N_{\rm W} R}\frac{\Delta R}{R} \tag{3}$$

where $N_{\rm W} = n\Delta R$ is the column density of the ionized material, so that

$$\frac{\Delta R}{R} = \frac{\xi N_{\rm W} R}{L} \tag{4}$$

This gives the lines A and B on figure 5 for the O VII and O VIII absorbing regions respectively.

We obtain additional constraints from the recombination timescales of the oxygen ions in the material. The recombination timescale (to all atomic levels) for highly ionized oxygen is given by

$$t_{\rm rec} \approx 200 n_9^{-1} T_5^{0.7}\,{\rm s} \tag{5}$$

where $n = 10^9 n_9\,{\rm cm}^{-3}$ and $T = 10^5 T_5\,{\rm K}$ is the temperature of the warm material (Shull & van Steenberg 1982). Using the definition of $\xi$, this can be expressed as

$$t_{\rm rec} \approx 200 \xi_2 R_{16}^2 L_{43}^{-1} T_5^{0.7}\,{\rm s} \tag{6}$$

where $\xi = 10^2 \xi_2\,{\rm erg\,cm\,s}^{-1}$, $R = 10^{16} R_{16}\,{\rm cm}$ and $L = 10^{43} L_{43}\,{\rm erg\,s}^{-1}$. The ASCA observations require that the O VIII absorber responds on timescales shorter than $10^4$ s. We take this to be the limit on the recombination timescale of the O IX ions from which the O VIII ions arise. Thus, assuming $L_{43} = 3$ and utilizing the measured value of $\xi$ for the O VIII absorber gives $R_{16} < 14$ (corresponding to a density limit of $n > 2 \times 10^7\,{\rm cm}^{-3}$). This is shown as limit on figure 5. Similarly, the constancy of the O VII edge implies a recombination timescale for O VII longer than $10^6$ s which gives $R_{16} > 300$ (corresponding to a density limit of $n < 2 \times 10^5\,{\rm cm}^{-3}$). This is limit D on figure 5.

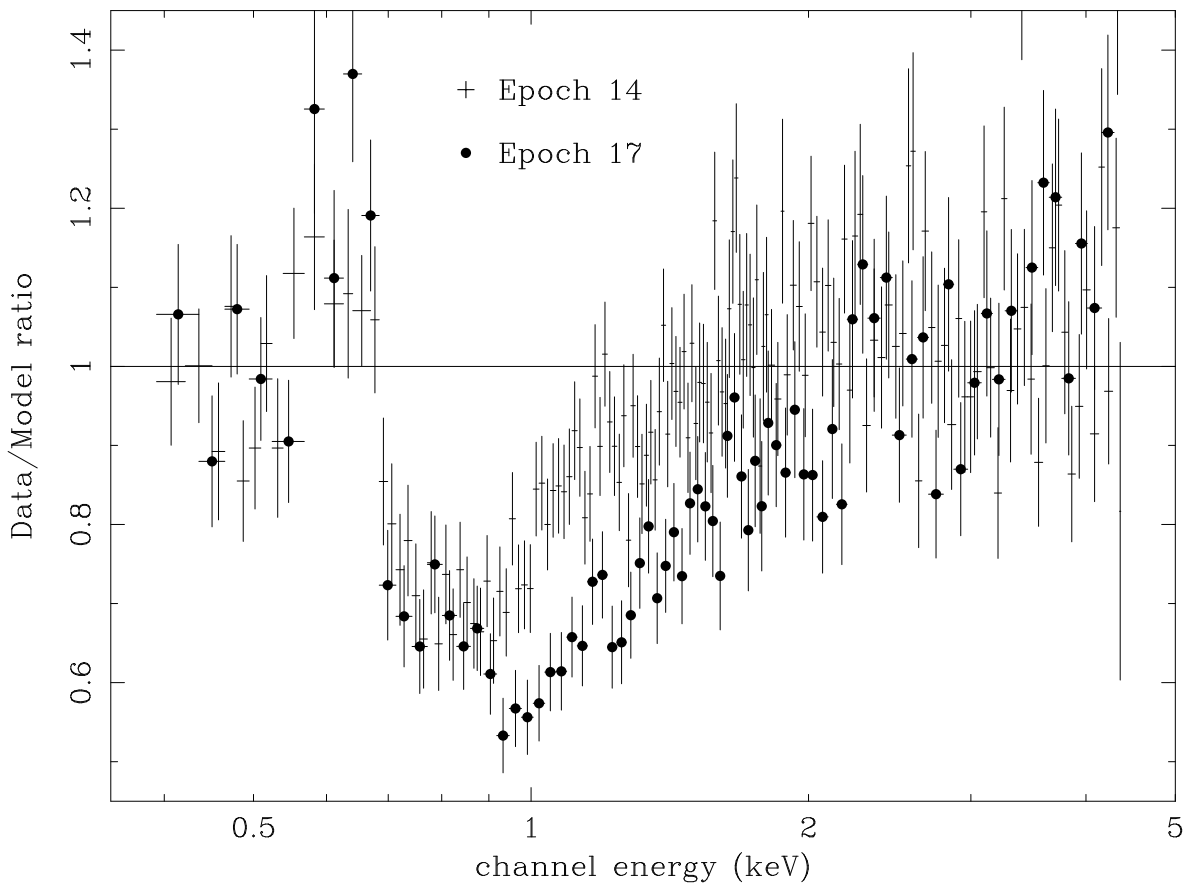

Fig. 3. The comparison of absorption edge features at epoch 14 and epoch 17 are shown. The vertical axis shows the ratio from the best fit power-law determined by the higher and lower energy continuum. The difference above 0.9 keV shows the difference of the depth of O VIII absorption edge.

Recombination to both O VIII and O VII leads to line emission at 0.65 and 0.57 keV, respectively. The effective fluorescent yield for this process is about 0.5, so significant lines are expected if the warm absorber has a high covering fraction. As mentioned already, there are large changes in the response function of the SIS near these energies so we must be very cautious in interpreting observations of such lines. Nevertheless, we do find that the observed flux that can be attributed to line emission is consistent with the above yield provided that the total covering fraction of the warm absorber is about one half.

### 3.2. *The physical nature of the warm material*

The constraints presented above demonstrate that the absorption plausibly occurs in two spatially distinct regions along the line of sight to the primary source. The O VIII absorber is constrained to be at radii characteristic of the broad line region (BLR) with a small volume filling factor ($\Delta R/R < 0.003$) and a density of $n > 2 \times 10^7\,{\rm cm}^{-3}$. Photoionization models give the tem-perature of this material as $T \sim 10^6$ K leading to a lower limit on the pressure of $nT > 2 \times 10^{13}$ cm$^{-3}$ K, similar to the pressure of the BLR. It is tempting to identify this with optically-thin clouds in the BLR which are in approximate pressure balance with the regular broad emission line clouds.

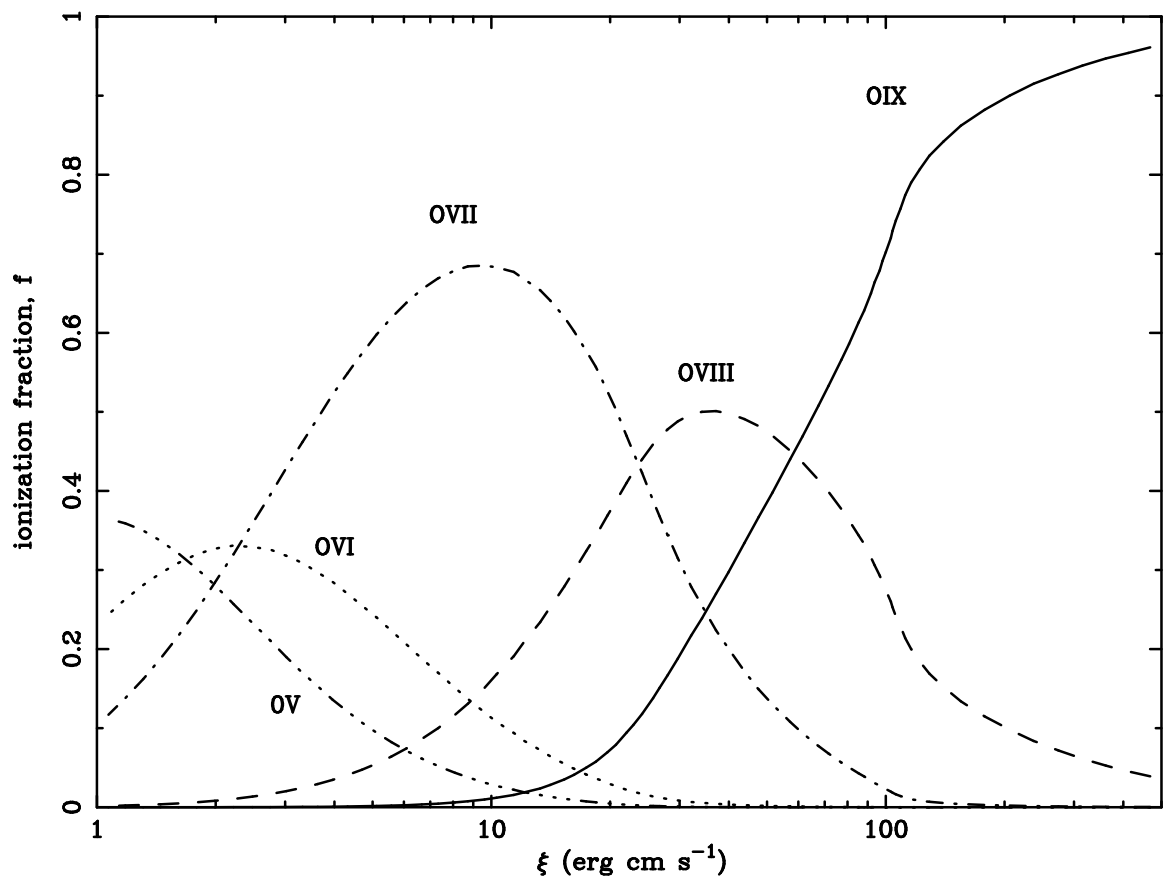

Fig. 4. Ion population of Oxygen for the ionization parameter $\xi$.

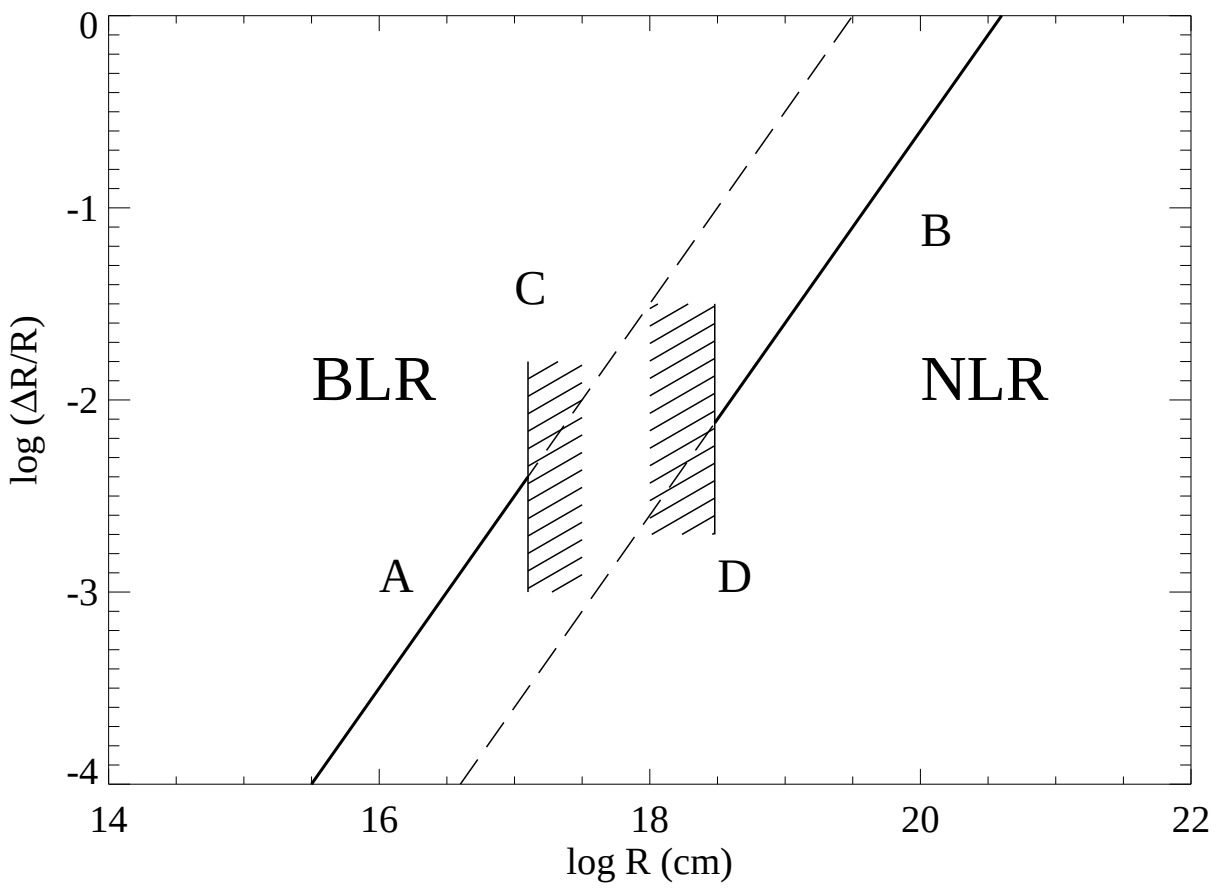

Fig. 5. The constraints in $R$–$(\Delta R/R)$ diagram for the inner and outer warm absorbers in the two warm absorber scheme. The lines A and B show the limits from equation (4) using the determined $\xi$ values in the warm absorber model. The limits C and D are obtained from equation (6) and the timescales of the column density changes of O VIII and O VII, respectively. The typical distances for BLR and NLR are also indicated.

We note that much of the oxygen in the inner region must be highly ionized (i.e. O IX) and so act as a reservoir for producing O VIII when the flux decreases. The column density of this medium may be $\sim 5 \times 10^{22}$ cm$^{-2}$ and it may be more space filling than the O VIII clouds. Just what its filling factor is can be estimated if it is assumed that the O IX material is heated by radiation and is in pressure equilibrium with the O VIII gas. The temperatures of the O IX and O VIII gas are then about $2 \times 10^5$ K and $10^6$ K, respectively (see photoionization plots in Reynolds & Fabian 1995). Consequently the density of the O IX gas is about 5 times smaller than that of the O VIII gas and its volume filling factor is less than 2 per cent (scaling from figure 5). The column density required for the O IX gas to be space filling is thus at least 50 times that of the O VIII absorber or about $8 \times 10^{23}$ cm$^{-2}$. The Thomson depth of such a medium then exceeds 50 per cent and the variability of the nucleus begins to be smeared out. The above values are taken for the limiting case; more reasonable values lead to the O IX medium being Thomson thick. Such results are contrary to the observed rapid X-ray variability seen in the source (note that electron scattering in the medium would also affect the shape of optical emission line profiles from BLR clouds). We conclude that the O IX reservoir which produces the O VIII absorber is cloudy and pressure-confined but some other medium (or even magnetic fields; Rees 1987) which is yet hotter and less dense.

The O VII absorber is significantly further from the central source and more volume filling ($\Delta R/R > 0.005$). It might be associated with a wind from the putative molecular torus and/or the scattering medium responsible for the scattered continuum in Seyfert 2 galaxies (such as in the models presented by Krolik & Kriss 1995). It should be stressed that it is implausible for the O VII edge not to vary unless the associated material is very diffuse (leading to a long recombination timescale). Figure 5 shows that there is no region of parameter space over which the fraction of O VII remains constant when the ionizing flux varies by factors of a few. Moreover, models in which an inner O VIII absorbing region shields an outer O VII absorbing region tend to *amplify* the observed variability of the O VII edge. To see this, consider a decrease in flux. Recombination of the O IX ions produce a deeper O VIII edge which removes photons capable of ionizing O VII. Consequently, the decrease in the ionizing flux at the location of the O VII absorbing region over-responds to the original decrease thereby producing an over response in the absorption edge. Strictly, this argument is only valid in the regime where $f_{O7}$ decreases with $\xi$. However, we know that the O VII absorber must exist in this regime due to the absence of an observed O VI edge or indeed oxygen L (and other) absorption (see figure 3).

It is interesting that the column densities of O VII and O VIII ions are comparable. Moreover, the absorption edges in MCG–6–30–15 and many other objects with warm absorbers have maximum optical depths which are of the order of unity. This could be due to the action of radiation pressure on the warm material. Such models require further development.

We are grateful to the all ASCA team members. CO is supported by Special Postdoctoral Researchers Program of The Institute of Physical and Chemical Research (RIKEN). ACF thanks the Royal Society for support.